\documentclass[twocolumn,noshowpacs,preprintnumbers,amsmath,amssymb]{revtex4}

\usepackage{graphicx}
\usepackage{dcolumn}
\usepackage{bm}
\usepackage{hyperref}

\begin{document}
\keywords{Graphene, fluorographene, graphite monofluoride, band gap, optical spectra, exciton, many-body theory.}
\title[Band Gaps and Optical Spectra from Single- and Double-Layer Fluorographene to Graphite Fluoride]{Band Gaps and Optical Spectra from Single- and Double-Layer Fluorographene to Graphite Fluoride: Many-Body Effects and Excitonic States}
\author{Franti\v{s}ek Karlick\'{y}}
\email{frantisek.karlicky@upol.cz}
\author{Michal Otyepka}
\email{michal.otyepka@upol.cz}
\affiliation{Regional Centre of Advanced Technologies and Materials, Department of Physical Chemistry, Faculty of Science, Palack\'{y} University, T\v{r}. 17. listopadu 12, 771 46 Olomouc, Czech Republic.}

\begin{abstract}
We compare first-principle band gaps and optical absorption spectra of single- and double-layer fluorographene with bulk graphite fluoride.
The electronic properties are calculated using the many-body GW approximation and the optical spectra using the Bethe-Salpeter equation (BSE). 
The inclusion of electron-hole interactions is crucial for predicting low energy excitonic absorption peaks. 
The position of the first exciton peak is identical in single-, double- and multilayer fluorographene, which may indicate that the onset of the absorption spectra does not differ in these materials. 
\end{abstract}

\maketitle


\section{Introduction}
Graphite fluoride, bulk (CF)$_n$, is a well-known graphite derivative
in which fluorine atoms are attached to carbon atoms by covalent C-F bonds.
Although bulk graphite fluoride has been used as a lubricant for nearly 100 years and has also been exploited as an excellent electrode material in primary lithium batteries, fluorographene (single-layer CF or graphene fluoride) was not prepared and isolated until 2010 \cite{nair10,zboril10}.
This reignited interest in the halogenation of sp$^2$-planar carbon materials, and halogenated graphenes were subsequently reported to have a plethora of remarkable properties \cite{karlicky13a} that are not observed in graphene itself.
The attachment of fluorine atoms changes the sp$^2$ hybridization state of the carbons in graphene to sp$^3$, which significantly affects the material's electronic properties and local structure.
These structural changes induce the opening of the band gap at the K point and lead to a loss of the $\pi$-conjugated electron cloud above and below the graphene plane.

Density functional theory (DFT) is a powerful tool for modeling graphene and its halogenated derivatives.
Because DFT methods based on the generalized gradient approximation (GGA) tend to underestimate the band gaps of graphene derivatives, the electronic properties of these materials are often described using
high-level GW many-body theory\cite{hedin65}.
However, there are apparent discrepancies between the band gaps predicted for fluorographene using the GW  approximation (GWA, 7-8 eV)\cite{klintenberg10,leenaerts10,samarakoon11,liang11,karlicky13,wei13} and the experimentally determined gap ($>$3 or $>$3.8 eV)\cite{nair10,jeon11}.
Such discrepancies are not necessarily problematic because calculated electronic band gaps should not be directly compared to the energies of electron transitions derived from optical spectra.
The electron transitions observed in optical spectra involve exciton formation 
and their energies can be estimated using the Bethe-Salpeter equation (BSE)\cite{bethe51}, which includes excitonic effects.
Fluorographene (CF) has a high exciton binding energy ($\sim$2 eV)\cite{karlicky13,wei13}, which may partially explain the discrepancy between the experimental data and the GWA band gap. The role of point defects in real materials was predicted to be synergistic but small\cite{karlicky13}.
The geometric structure of graphite monofluoride, i.e. bulk (CF)$_n$, has been studied extensively because it was not determined satisfactorily for some time; it is only quite recently that its structure
was unambiguously resolved (see Refs. \cite{sato04,wang10,fujimoto97,touhara87} and references therein). In contrast, there have been comparatively few detailed studies on the electronic and optical properties of (CF)$_n$ (see e.g. Refs. \cite{cheng10,wang10}).

The aim of this work was to determine how the band gaps and absorption spectra of single- and multi-layer fluorographenes depend on the number of layers in the material. 
The electronic and optical properties of the studied materials were modeled using the GW plus Bethe-Salpeter equation approach, which accounts for both electron-electron and electron-hole correlation effects.
Particular attention was paid to the convergence of the GW and BSE calculations in order to obtain highly accurate estimates of both band gaps and optical transitions.
 

\section{Methods}
The projector augmented waves (PAW) method \cite{blochl94} as implemented in the Vienna ab initio simulation package (VASP)\cite{kresse99} was used to perform total energy and absorption spectra calculations on structures generated by geometry optimization using the Perdew-Burke-Ernzerhof (PBE) GGA functional \cite{perdew96}.
Geometry optimizations were also performed using the PBE-D2 functional \cite{grimme06} and non-local vdW functionals available in VASP \cite{rem_functionals}. 
The unit cell was obtained by minimizing the total energy as a function of the lattice parameter. 
For each value of the lattice constant, atomic positions (i.e. internal degrees of freedom) were relaxed until the change in forces on each atom was less than $1\times 10^{-3}$ eV/\AA \, (the break condition for the electronic step was an energy difference of $1\times 10^{-6}$ eV). 
$20\times 20\times 1$ and $20\times 20\times 3$ {\bf{k}}-points (including the $\Gamma $ point) sampled the first Brillouin zone of 2D and 3D materials, respectively. 
As periodic boundary conditions were applied in all three dimensions, distances of 30 and 36 \AA \, between repeated cells in the out-of-plane direction were considered for CF and (CF)$_2$, respectively, to minimize (spurious) interactions between adjacent layers. 
A cut-off energy of 500 eV was applied for the plane-wave basis set.

The quasi-particle (QP) energies $E_{n\bf{k}}^{\mathrm{QP}}$ within the GWA were calculated as first-order corrections to the DFT single-particle energies $E_{n\bf{k}}$
\begin{equation}
\label{G0W0}
E_{n\bf{k}}^{\mathrm{QP,0}}=Re[\langle \phi_{n\bf{k}} \lvert T+V_{\mathrm{n-e}}+V_{\mathrm{H}}+\Sigma_{xc}(G,W;E_{n\bf{k}}) \rvert \phi_{n\bf{k}} \rangle]
\end{equation}
where $T$ is the kinetic energy operator, $V_{\mathrm{n-e}}$ the potential of the nuclei, $V_{\mathrm{H}}$ the Hartree potential, and $n$ and $\bf{k}$ are the band and $\bf{k}$-point indices, respectively. 
Both Green's function $G$ and the screened Coulomb interaction $W$ in the self-energy operator $\Sigma$ were calculated using DFT single-particle energies and wave functions
generated using either the GGA PBE functional or the screened hybrid functional of Heyd-Scuseria-Ernzerhof (HSE06)\cite{krukau06}.  
The corresponding single-shot total energies are henceforth referred to as G$_0$W$_0$@PBE and G$_0$W$_0$@HSE06.
We also obtained an updated quasiparticle energy $E_{n\bf{k}}^{\mathrm{QP,1}}$ by performing a single iteration in $G$ ($W$ was fixed at the initial HSE06 value $W_0$). The resulting total energy was 
designated G$_1$W$_0$@HSE06.

Excitonic effects were accounted for using the Bethe-Salpeter equation \cite{bethe51}, which corresponds to the following excitonic equation \cite{rohlfing00}:  
\begin{equation}
\label{BSE}
(E_{c\bf{k}}^{QP}-E_{v\bf{k}}^{QP}) A_{vc\bf{k}}^S + \sum_{v'c'\bf{k}'} \langle {vc\bf{k}} \lvert K_{\mathrm{e-h}} \rvert {v'c'\bf{k}'} \rangle = \Omega^S A_{vc\bf{k}}^S
\end{equation}
This method is also implemented in VASP\cite{fuchs08}. 
The $A_{vc\bf{k}}^S$ values obtained by diagonalization of Eq. \eqref{BSE} are the amplitudes of a free electron-hole pair configuration composed of the electron state $\rvert c\bf{k}\rangle$ and the hole state $\rvert v\bf{k}\rangle$, the $\Omega^S$ values are eigenenergies (corresponding to the exciton excitation energies), and $K_{\mathrm{e-h}}$ is the electron-hole interaction kernel.
 
The optical absorption spectrum corresponded to the imaginary part of the dielectric function $\varepsilon(\omega )$. We considered only light absorption with polarization in the plane defined by the CF layer. A gaussian broadening of 50 meV and 512 sampling points were used for the dielectric function. 


\section{Results and Discussion}

\subsection{Geometrical Structure and DFT Band Gaps}

There have been many discussions concerning the structure of graphite fluoride (CF)$_n$ since its first synthesis by Ruff {\em{et al.}} in 1934\cite{ruff34}.
This is largely because it was very difficult to prepare a single crystal of graphite fluoride, and only limited information could be obtained from powder diffraction analysis of the polycrystal.
The hexagonal lattice parameters for (CF)$_n$ are slightly dependent on the crystallinity of the starting carbon materials and the fluorination temperature\cite{touhara87}.
Moreover, the fluorinated carbon layers of (CF)$_n$ are randomly stacked, so the precise determination of the structure parameters of these compounds using the conventional Bragg diffraction patterns was quite difficult.

The lattice parameters reported in the literature for (CF)$_n$ vary widely. We therefore briefly review and discuss the experimental data on this material that have been gathered over the last 80 years.
The reader is referred to the literature (e.g. the introduction of Ref. \cite{fujimoto97}) for an overview of the early data.
Nowadays, (CF)$_n$ is generally accepted to exist in the chair-type structure rather than the boat-type alternative\cite{touhara87,fujimoto97,sato04,cheng10,charlier93,han10}. 
In 1987, Touhara {\em{et al.}}\cite{touhara87} took Laue photographs of X-ray diffraction (XRD) from (CF)$_n$ synthesized from natural graphite and highly oriented pyrolytic graphite (HOPG). 
They prepared a highly stoichiometric material (F/C ratio = 1.0), determined it to have a 6-fold symmetric structure, and concluded that it exhibited AA' stacking ($P\bar{6}m2$ space group) with lattice constants $a$ = 2.53-2.57\AA, and $c$ = 12.1 \AA \,(i.e. $c$ = 6.05 \AA \, corresponding to classical AA stacking). 
Fujimoto \cite{fujimoto97} reported a slight preference for the AA stacking sequence ($P3m1$ space group) in 1997 and confirmed Touhara's lattice constants by performing Rietveld analysis on mesocarbon microbeads with an average diameter of $\sim6 \mu$m. 
However, he also stated that the stacking sequence can depend on the carbon material used to prepare the (CF)$_n$ and the fluorination conditions, and that individual samples can contain several structures and stacking types.
Sato {\em{et al.}} \cite{sato04} obtained lattice constants of $a$ = 2.58 \AA \, and $c$ = 6.23 \AA \, using XRD in 2004. Relatively large average crystallite sizes were achieved along the $c$- and $a$-axes ($L_c$ = 6.1 nm and $L_a$ = 58 nm), albeit with an F/C ratio of 1.19. 
These authors then used neutron diffraction data to refine the XRD structure and obtained an accurate lattice constant of $a$ = 2.60-2.61 \AA, along with C-C and C-F bond lengths of $r_{\mathrm{C-C}}$ = 1.58 \AA \, and $r_{\mathrm{C-F}}$ = 1.36 \AA \, and bond angles; see Table \ref{tab_geom} for all parameters.
Recently, Cheng {\em{et al.}}\cite{cheng10} reported lattice constants of $a$ = 2.57\AA \, and $c$ = 6.2 \AA \, based on XRD analyses of SP-1 graphite with an F/C ratio of $\sim$ 1. The same group (Wang {\em{et al.}} \cite{wang10}) subsequently determined a value of $c$ = 5.82 \AA \, based on the XRD spectrum of fluorinated HOPG (F/C = 0.96 with a small proportion of >CF$_2$ bonds, $L_c \sim$ 6 nm, $L_a \sim$ 20 nm).

\begingroup
\squeezetable
\begin{table}[t]
\caption{\label{tab_geom}Geometrical parameters (lattice constants {\em{a}} and {\em{c}} and bond lengths {\em{r}} in \AA \, and bond angles in degrees) and DFT band gaps (in eV) of bulk (CF)$_n$ obtained with various functionals \cite{perdew96,rem_functionals,grimme06} and by experimental measurement.}
\begin{ruledtabular}
\begin{tabular}{lccccccc} 
Method&\em{a}&\em{c}&{\em{r}}$_{\mathrm{C-C}}$&{\em{r}}$_{\mathrm{C-F}}$&$\angle$C-C-C&$\angle$F-C-C&{\em{E}}$_{\mathrm{gap}}$\\
\hline
PBE-D2&2.60&5.71&1.58&1.38&110.9&108.0&3.51\\
vdW-DF&2.62&5.79&1.59&1.40&111.2&107.7&3.71\\
vdW-DF2&2.62&5.70&1.59&1.41&111.3&107.5&3.78\\
optB88-vdW&2.59&5.66&1.57&1.38&110.9&108.0&3.57\\
optPBE-vdW&2.61&5.70&1.58&1.39&111.0&108.0&3.65\\
PBE& 2.60 & 6.08&1.58&1.38&110.9&108.0&3.38 \\
exp.\cite{sato04}&2.60-2.61&6.23&1.58&1.36&111&108& \\
exp.\cite{touhara87,fujimoto97}&2.53-2.57&6.05& & & & & \\
exp.\cite{cheng10}&2.57&6.2& & & & & \\
exp.\cite{wang10}& &5.82& & & & &$\sim $5 \\
\end{tabular}
\end{ruledtabular}
\end{table}
\endgroup
 
All of the species considered in our calculations were assumed to adopt hexagonal chair-like conformations because computational results indicate that this is the most stable conformation of CF \cite{leenaerts10} and (CF)$_n$\cite{charlier93,han10}.
For (CF)$_n$ and (CF)$_2$ we used AA stacking ($P3m1$ space group) because this is the structure most commonly reported for experimental samples of bulk (CF)$_n$, as discussed above.
The CF layers in (CF)$_n$ are weakly coupled by van der Waals (vdW) forces. 
We therefore optimized the structure of (CF)$_n$ using several DFT functionals that account for vdW interactions (see Table \ref{tab_geom}) in addition to the standard PBE functional.
One might expect a correct treatment of vdW forces to be crucial for obtaining reliable values of the $c$ lattice parameter because GGA PBE does not account for non-local electron correlation effects and fails for vdW interactions\cite{lazar13}. 
As shown in Table \ref{tab_geom}, all of the tested DFT functionals predicted $a$ lattice parameters, bond lengths and angles in perfect agreement with the neutron diffraction data from Ref. \cite{sato04}.
On the other hand, all of the DFT functionals that account for non-local correlation effects yielded $c$ lattice parameters in the range of 5.66-5.79 \AA, which is just below the range determined 
experimentally ($c$ = 5.8-6.2 \AA). 
The PBE functional performed surprisingly well, giving a $c$ lattice parameter of 6.08 \AA, in good agreement with the experimental data. 
These findings indicate that current DFT functionals that include non-local correlation effects may overestimate the strength of the interactions between individual graphene layers.
As discussed above, the $c$ lattice parameter depends on the F/C ratio and the nature of the carbon starting material. Taking these factors into account, the $c$ values of 6.05 \AA \, and 6.2 \AA \, from Refs. \cite{touhara87,fujimoto97} and \cite{cheng10} are likely to be the most relevant for our purposes because they were determined for materials with F/C ratios of $\sim$ 1.
We therefore used the geometric structure for (CF)$_n$ obtained by PBE optimization (which had a $c$ value of 6.08 \AA ) in our many-body calculations. This interlayer distance was also used when performing
calculations for (CF)$_2$
For CF, we used geometrical parameters obtained by PBE optimization in our previous works\cite{karlicky13,karlicky12}, which are in agreement with the PBE parameters for (CF)$_n$ (Table \ref{tab_geom}).

The DFT band gap of (CF)$_n$ was larger than that for CF: using the PBE functional, the gaps for the two materials are 3.38 eV and 3.09 eV\cite{karlicky13}, respectively. 
The band gap of (CF)$_n$ varied with the functional used (Table \ref{tab_geom}).
This functional dependence was partly due to the band gap's slight sensitivity to the $c$ lattice parameter: the
PBE band gaps obtained using fixed $c$ values of 5.72 \AA \, and 6.19\AA \, (roughly corresponding to the two extremes of the $c$ value range from Table \ref{tab_geom}) differed by 0.1 eV.
The functional dependence was also  
partly due to the different geometric parameters obtained for single CF layers with the tested functionals (Table \ref{tab_geom}) and the differences between the functionals' exchange components. 
Significantly larger band gaps could be obtained using hybrid density functionals (see Ref. \cite{karlicky12} and Sec. 3.3). 
We note that the electronic structures of layered materials are much more sensitive to the $a$ lattice parameter than the $c$ lattice parameter; it has therefore been suggested that in-plane strain could be used for band gap engineering in 2D materials\cite{conley13}. 



\subsection{Convergence Parameters for Many-Body Methods}

To obtain accurate QP band gaps and reliable optical spectra, it is essential to perform a convergence study covering several parameters (such as the size of the {\bf{k}}-point grid).
In addition, some computational difficulties with certain parameters have recently been reported for various materials.
GW gap calculations were found to exhibit false convergence with respect to the number of bands, $N_\mathrm{b}$, in studies on single-layered MoS$_2$ \cite{qiu13}and bulk ZnO \cite{shih10} when using unconverged energy cutoffs for the dielectric matrix, $E_\mathrm{cut}^\mathrm{GW}$.
To obtain a band gap for MoS$_2$ with an accuracy better than 0.1 eV, it was necessary to employ a significantly higher $E_\mathrm{cut}^\mathrm{GW}$ value of 476 eV than was reported for previous calculations and also to include a greater number of bands ($N_\mathrm{b}$ = 16000).
It was also essential to use an appropriate number of {\bf{k}}-points to obtain accurate predicted absorption spectra for MoS$_2$\cite{qiu13}.
We therefore decided to investigate these parameters carefully before determining final band gaps and spectra for CF and its stacked derivatives.

\begin{figure}
  \includegraphics[width=\columnwidth]{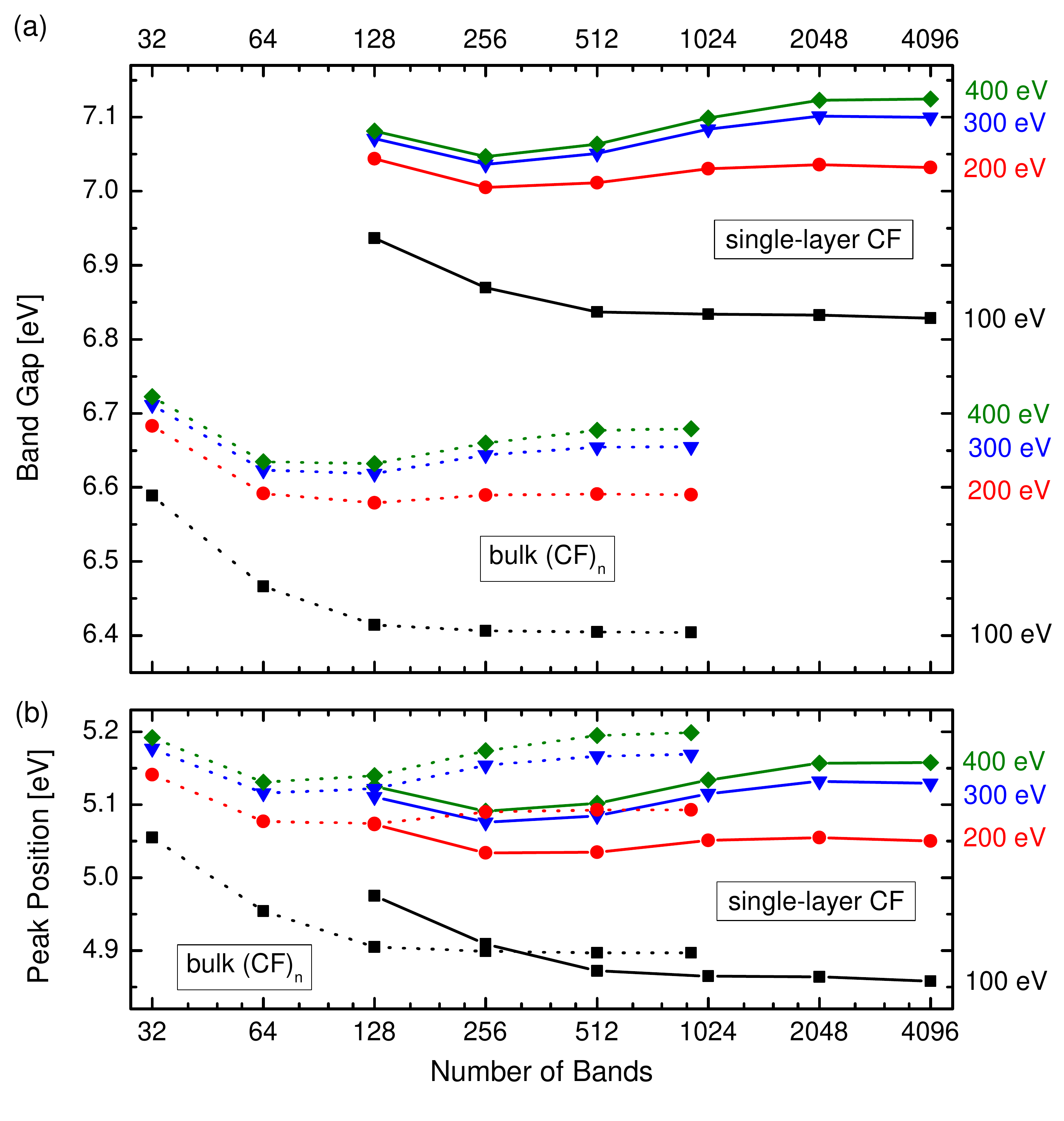}%
  \caption{\label{fig_b_conv} 
(a) Quasiparticle band gaps of single-layer CF and bulk (CF)$_n$ calculated using the G$_0$W$_0$@PBE method with different dielectric matrix cutoffs showing the false convergence behavior of the band gap when a small cutoff energy is used. $12\times 12\times 1$ and $12\times 12\times 3$ {\bf{k}}-point grids were used for CF and (CF)$_n$, respectively. (b) The first exciton peak exhibited similar false convergence behavior to the band gap.}
\end{figure}

The GW band gap, $E_\mathrm{gap}^\mathrm{GW}$, converged quite quickly with the number of {\bf{k}}-points and generally converged better than the calculated optical spectra (see below). 
A $12\times 12 \times 3$ grid gave $E_\mathrm{gap}^\mathrm{GW}$ convergence within 0.01 eV.
Using this {\bf{k}}-point grid, we tested the convergence achieved with different values of $N_\mathrm{b}$ and $E_\mathrm{cut}^\mathrm{GW}$ for CF and (CF)$_n$, as shown in Figure \ref{fig_b_conv}a. 
A dielectric matrix energy cutoff of at least $E_\mathrm{cut}^\mathrm{GW}$ = 300-400 eV was required to achieve reliable convergence.
In addition, it was necessary to include several hundred bands to achieve convergence for bulk (CF)$_n$ and at least one thousand bands for single-layer CF.
The density of bands was much greater for CF than for (CF)$_n$. For example, CF had $\sim$530 bands at energies below 150 eV whereas (CF)$_n$ had $\sim$140.
We ultimately selected $N_\mathrm{b}$ values of 256 and 1024 for (CF)$_n$ and CF, respectively, and an $E_\mathrm{cut}^\mathrm{GW}$ value of 300 eV for both materials.
Converged valence and conductance bands with accuracies better than 0.05 eV were obtained using these parameters.
Figure \ref{fig_b_conv}a clearly shows that attempts to converge $N_\mathrm{b}$ with an unconverged $E_\mathrm{cut}^\mathrm{GW}$ value may lead to false convergence behaviour\cite{shih10}.
The band gap error due to the use of an unconverged $E_\mathrm{cut}^\mathrm{GW}$ value was $\sim$0.2 eV ($E_\mathrm{cut}^\mathrm{GW}$ = 400 eV $\rightarrow$ 100 eV), which is similar to that reported 
for MoS$_2$\cite{qiu13}. However, the convergence of the band gap with respect to $N_\mathrm{b}$ for MoS$_2$ was much slower than for CF, and errors of up to 1 eV were observed for the former material.

Figure \ref{fig_b_conv}b shows that similar convergence behavior with respect to $E_\mathrm{cut}^\mathrm{GW}$ and $N_\mathrm{b}$ was observed for the first exciton peak of single-layer and bulk CF. The trend was same as for the band gap $E_\mathrm{gap}^\mathrm{GW}$, implying that there is some cancellation of errors when calculating the exciton binding energy $E_\mathrm{b}$. All of the exciton binding energies obtained using various combinations of $E_\mathrm{cut}^\mathrm{GW}$ and $N_\mathrm{b}$ in Figure \ref{fig_b_conv} were between 1.48-1.54 eV for bulk (CF)$_n$ and between 1.93-1.97 for single-layer CF.
That is to say, the fortuitous cancellation of errors meant that reasonable values of the exciton energy were obtained even from calculations that were not fully converged.

\begin{figure}
  \includegraphics[width=\columnwidth]{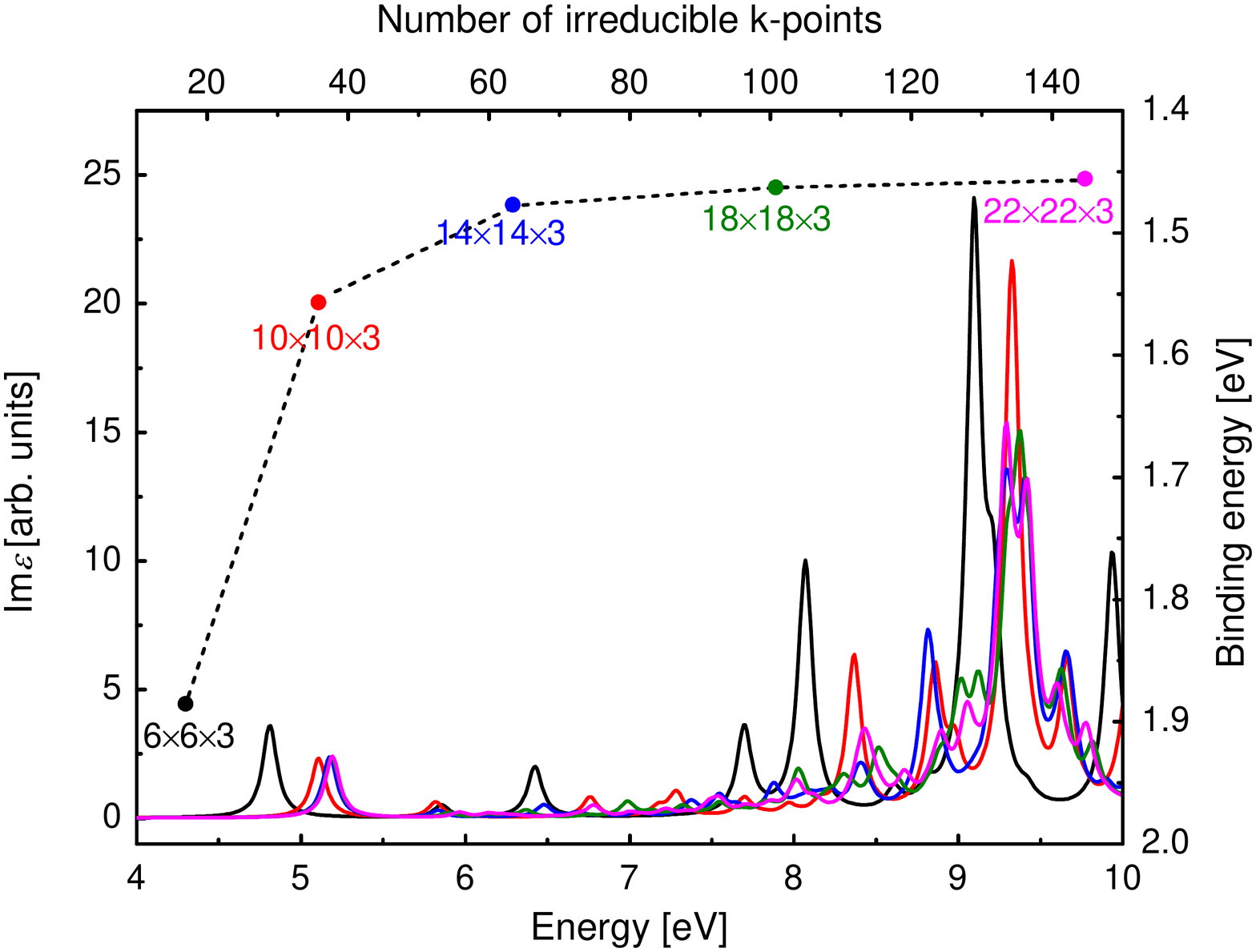}%
  \caption{\label{fig_k_conv} 
Left axis: BSE+G$_0$W$_0$@PBE spectra of bulk (CF)$_n$ for different {\bf{k}}-point grids. Right axis: Exciton binding energy as a function of the number of irreducible {\bf{k}}-points (top axis). Note the correspondence between the colors of the spectra and the binding energies.}
\end{figure}

In order to obtain accurate absorption spectra, it is important to pay attention to the convergence properties of the Bethe-Salpeter equation. 
There are two important factors that affect the BSE Hamiltonian: the number of Bloch bands and the number of {\bf{k}}-points.
We included all valence (11) and 13 conduction bands for CF and (CF)$_n$, and 14 valence and 14 conduction bands for (CF)$_2$ (these values correspond to the maximal values taken by the indices $v$ and $c$ in Equation \eqref{BSE}) to obtain the dielectric function $\varepsilon(\omega )$ because we were only interested in the lower energy region of the spectrum.
However, we found that the positions of the exciton peaks were very sensitive to the number of  {\bf{k}}-points - see Figure \ref{fig_k_conv} for (CF)$_n$ and Figure 4a from Ref. \cite{karlicky13} for CF. 
It was necessary to use at least a $14\times 14\times 3$ grid for (CF)$_n$ and an $18\times 18$ grid for CF in order to achieve exciton binding energy ($E_\mathrm{b}$) convergence to within 0.02 eV.
We therefore decided to use a $20\times 20$ {\bf{k}}-point grid when evaluating the electron-hole kernel in the BSE for all considered materials. 
The dominant exciton peak at 9-10 eV was not especially well converged with this grid, but the use of denser grids would have imposed extreme demands in terms of memory and CPU time for all of the studied
materials other than bulk CF.

The difference in the behavior of these excitons can also be understood by considering their wave functions. 
An $n\times n$ grid in reciprocal space can only be used to map an exciton wave function within a real-space $n\times n$ supercell. Therefore, an unduly small grid will impose an artificial confinement on the exciton and thus increase its predicted binding energy. 
The first exciton wave function of CF had a relatively small radius, covering a volume approximately equivalent to a $7\times 7$ supercell in real space\cite{wei13}.
However, the wave function of the dominant peak at 9-10 eV was more spread out and therefore converged more slowly with increasing grid size. 


\subsection{Optical Spectra}

\begin{figure}
  \includegraphics[width=\columnwidth]{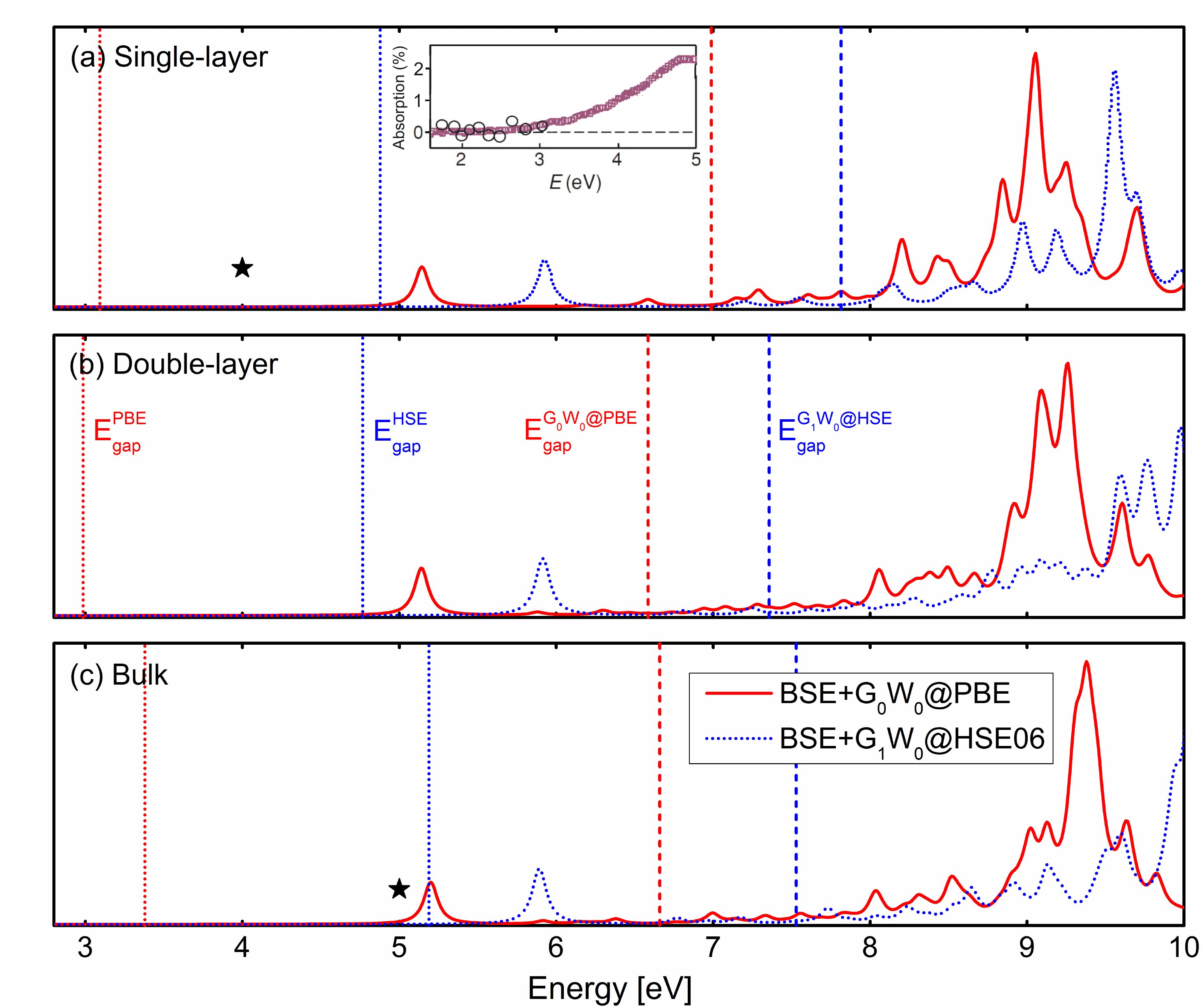}%
  \caption{\label{fig_spectra} 
Optical spectra for (a) single-layer, (b) double-layer, and (c) bulk CF obtained with BSE+G$_0$W$_0$@PBE and BSE+G$_1$W$_0$@HSE06. DFT and GW band gaps at the $\Gamma$ point are indicated by vertical lines. The experimental optical spectrum\cite{nair10} for single-layer CF is shown in the inset of panel (a) for comparative purposes. Estimated experimental values for the optical band gaps of single-layer and bulk CF \cite{jeon11,wang10} are indicated by asterisks.}
\end{figure}

\begin{table}[t]
 \caption{\label{tab_energ}DFT band gaps $E_\mathrm{gap}$, QP band gaps $E_\mathrm{gap}^\mathrm{GW}$, excitation energies $E_\mathrm{exc}^\mathrm{BSE}$, and exciton binding energies $E_\mathrm{b}$ (in eV) of single-layer, double-layer, and bulk CF.}
\begin{ruledtabular}
\begin{tabular}{lccc}
 &CF&(CF)$_2$&(CF)$_n$\\
\hline
$E_\mathrm{gap}^\mathrm{PBE}$&3.09&2.99&3.38\\
$E_\mathrm{gap}^\mathrm{HSE06}$&4.88&4.77&5.19\\
$E_\mathrm{gap}^\mathrm{G_0W_0@PBE}$&6.99&6.59&6.66\\
$E_\mathrm{gap}^\mathrm{G_1W_0@HSE06}$&7.82&7.38&7.53\\
$E_\mathrm{exc}^\mathrm{BSE+G_0W_0@PBE}$&5.14&5.15&5.20\\
$E_\mathrm{exc}^\mathrm{BSE+G_1W_0@HSE06}$&5.92&5.91&5.89\\
$E_\mathrm{exc}^\mathrm{exp.}$&$>$3.8\cite{jeon11}&-&$\sim $5\cite{wang10}\\
$E_\mathrm{b}^\mathrm{BSE+G_0W_0@PBE}$&1.84&1.44&1.46\\
$E_\mathrm{b}^\mathrm{BSE+G_1W_0@HSE06}$&1.89&1.44&1.64\\
\end{tabular}
\end{ruledtabular}
\end{table}

Finally, optical absorption spectra for light polarization parallel to the surface plane are shown in Figure \ref{fig_spectra} for CF, (CF)$_2$, and (CF)$_n$.
Electron-hole excitations, which are not considered in GW theory itself, were clearly responsible for the optical absorptions at energies lower than the GW band gaps.
The main features of the spectra for CF, (CF)$_2$, and (CF)$_n$ were similar.
There was a cancellation effect between the band gap opening due to QP corrections and the red-shift of the optical absorption due to excitonic effects (i.e. the effects of electron-electron and electron-hole interactions were opposed).
The first exciton was degenerate with an exciton of similar oscillator strength, both of which were optically active (bright). 
These corresponded to dipole-allowed vertical transitions from the two highest occupied bands to the lowest empty band\cite{karlicky13,wei13}. 
Two other degenerate excitons were identified for (CF)$_2$; the corresponding peak was shifted by only 0.03 eV (to a higher energy) relative to the first exciton peak.
The position of the first exciton peak $E_\mathrm{exc}$ was remarkably stable with respect to the number of layers ($\sim $5.2 and $\sim $5.9 eV at the BSE+G$_0$W$_0$@PBE and BSE+G$_1$W$_0$@HSE06 levels, respectively; {\it{cf}}. Table \ref{tab_energ}).
On the other hand, the binding energy of the first exciton $E_\mathrm{b}=E_\mathrm{gap}^\mathrm{GW}-E_\mathrm{exc}$ varied with the number of layers due to differences in the QP band gap $E_\mathrm{gap}^\mathrm{GW}$ (Table \ref{tab_energ}). 
We note that similar trends were observed for both the PBE and HSE06 band gaps as can be seen from Figure \ref{fig_spectra} and Table \ref{tab_energ}.

The high energy regions (around 9-10 eV) of the optical spectra were dominated by electron-hole transitions originating from the two highest occupied bands and the
four lowest empty bands\cite{karlicky13}. 
The position corresponding to the high-energy peak was systematically shifted to higher energies on going from single-layer CF to double-layer and then bulk material; this may be useul for spectroscopic discrimination between samples with different layer numbers.

CF was found to be transparent at visible frequencies and to only start absorbing light in the blue range\cite{nair10}. It is anticipated to have a band gap of > 3.8 eV \cite{jeon11}. 
(CF)$_n$ appears white; photoluminescence studies suggest that its band gap is close to 5 eV \cite{wang10}. 
These values were significantly red-shifted with respect to the G$_0$W$_0$@PBE and G$_1$W$_0$@HSE06 band gaps ($\sim$ 7-8 eV).
The large exciton binding energies of CF and (CF)$_n$ therefore explain the optical band gaps obtained from photoluminescence experiments\cite{jeon11,wang10}.

We note that the optical absorption spectrum for light polarization perpendicular to the surface plane was almost negligible up to 9.5 eV. This indicates
that (CF)$_{1,2,n}$ exhibits strong optical anisotropy because completely different absorption spectra were obtained for light polarization parallel to the surface plane.

As regards the BSE+G$_0$W$_0$@PBE and BSE+G$_1$W$_0$@HSE06 methods, the latter should be the better from a theoretical perspective.
One self-consistent update of $G$ increases the gap substantially, to its fully $G$-self-consistent limit \cite{qiu13,karlicky13}: $E_\mathrm{gap}^\mathrm{G_1W_0} \sim E_\mathrm{gap}^\mathrm{G_{\infty}W_0}$.
In addition, $E_\mathrm{gap}^\mathrm{G_{\infty}W_0@HSE06}$ is close to the full $E_\mathrm{gap}^\mathrm{G_{\infty}W_{\infty}@PBE}$ \cite{karlicky13,shishkin07}.
On the other hand, comparisons with the available experimental data (Figure \ref{fig_spectra}, Table \ref{tab_energ}) clearly show that the single-shot BSE+G$_0$W$_0$@PBE method offers very good performance.
This is consistent with the established consensus for 3D materials: $E_\mathrm{gap}^\mathrm{G_0W_0}$ or $E_\mathrm{gap}^\mathrm{GW_0}$ provide the best
agreement with experimental band gaps, while $E_\mathrm{gap}^\mathrm{GW}$ often overestimates them\cite{shishkin07}. 


\section{Conclusions}
We have performed first-principles calculations of the quasiparticle band structure and excitonic properties of CF, (CF)$_2$ and (CF)$_n$ with many-body effects included. 
Reliable predicted band gaps and optical transitions could only be obtained by using high numbers of bands and {\bf{k}}-points, and carefully converged energy cutoffs.
The inclusion of electron-hole interactions was found to be crucial for predicting low energy excitonic absorption peaks. 
The position of the first exciton peak ($\sim$ 5.2 eV at G$_0$W$_0$@PBE level) was independent of the number of layers but the dominant high-energy peak was systematically shifted to higher energies on going from single- to double-layer CF and then to the bulk material ($\sim$ 9 $\rightarrow$ 9.5 eV). This may be useful for distinguishing between samples with different numbers of layers by spectroscopy. 
The large exciton binding energies for CF, (CF)$_2$ and (CF)$_n$ ($E_\mathrm{b}$ = 1.84, 1.44, and 1.46 eV, respectively) explained the optical band gaps determined by photoluminescence measurements. 

\section{Acknowledgments}
Financial support from the Czech Science Foundation (P208/12/G016), 
the Operational Program Research and Development for Innovations - European Regional Development Fund (project CZ.1.05/2.1.00/03.0058) and 
the Operational Program Education for Competitiveness - European Social Fund (project CZ.1.07/2.3.00/20.0017) 
is gratefully acknowledged.

\bibliography{andp} 

\end{document}